\documentstyle[12pt]{article}

\begin{document}
\vspace{2cm}

\begin{center}
{\Large {\bf {Statistical Properties 
and Algebraic Characteristics of Quantum Superpositions of Negative
Binomial States }}}
\end{center}

\vspace{1cm}

\begin{center}
{\large {\bf {\ Xiao-Guang Wang $^{\dagger }$\footnote{{\large {\bf {\bf %
E-mail: xyw@aphy.iphy.ac.cn}}}} and Hong-Chen Fu $^{*}$}}}
\end{center}

\vspace{0.5cm}

\begin{center}
{$\dagger $ 
CCAST(World Lab.), P.O.Box 8730, Beijing 100080\\
and\\
Laboratory of optical physics, Institute of Physics,
Chinese Academy of Sciences, Beijing 100080, P.R. China }

{* Institute for Theoretical Physics, Northeast Normal University, Changchun
130024, P.R.China }
\end{center}

\vspace{2cm}

\begin{center}
{\bf Abstract}
\end{center}

We introduce new kinds of states of quantized radiation fields, which are
the superpositions of negative binomial states. They exhibit remarkable
non-classical properties and reduce to Schr\"odinger cat states in a certain
limit. The algebras involved in the even and odd negative binomial states turn
out to be generally deformed oscillator algebras. It is found that the
even and odd negative binomial states satisfy a same eigenvalue equation with a same
eigenvalue and they can be viewed as two-photon nonlinear coherent states.
 Two methods of generating such states are proposed.

\newpage

\section{Introduction}

The binomial states introduced by Stoler in 1985[1] have attracted
much attention [2-5]. An important feature of the BS is that it interpolates between
the number state and the coherent state. The notion of of the BSs was also generalized to
the number-phase states[6],hypergeometric states[7] as well as
their q-deformation[8]. Sebawe Abdalla et al. introduced the
even binomial states [9] which interpolate between the even coherent states and
the even number states. Vidiella-Barranco et al. introduced the
superpositions of binomial states [10] which interpolate between the
so-called Schr\"{o}dinger cat states [11-15] and the number states.

Other interesting states are negative binomial states(NBSs)[16-22]. The NBS
is defined in the number state basis as 
\begin{equation}
|\eta _C;M\rangle=(1-\eta ^2)^{\frac M2}\sum_{n=0}^\infty {M+n-1\choose
n}^{1/2}(\eta _C)^n|n\rangle ,
\end{equation}
where $\eta _C=\eta \exp (i\theta )$, $M$ is a fixed positive integer , $%
\eta ^2$ is the probability satisfying $0<\eta ^2<1$, and $|n\rangle$ is the usual number(Fock) state. The photon number
distribution of the NBS is the negative
binomial probability distribution in classical probability theory [23]. Matsuo has
discussed the Glauber-Sudarshan $P$ representation of the NBS [16] and this
work is extended to obtain the quasiprobability distributions [17]. The
methods of generation of the NBS are proposed by Agarwal [19] and Fu [20].
Non-classical properties of the NBS and their interaction with two-level
atoms[21] have been investigated. Some statistical properties of 
even and odd NBSs are also discussed[22]. An interesting aspect of the NBS is
that it is an intermediate phase-coherent state in the sense that it
reduces to the Susskin-Glogower phase state[24] and the coherent state in two
two different limits.

In this paper, we introduce and study a new class of states, the quantum
superpositions of the NBSs. These states reduce to Schr\"odinger cat states in
a certain limit. The non-classical properties and algebraic characteristics of these states are studied in
detail. Two methods of generation of them are proposed.


\section{Normalization of the superpositions of the NBSs}


Now we define a superposition of two NBSs
\begin{equation}
|\phi ,\eta _C,M\rangle ={\cal N}\left[ |\eta _C,M\rangle +\exp
(i\phi )|-\eta _C,M\rangle \right]
\end{equation}
where $0\leq \phi \leq 2\pi $ and ${\cal N}$ is normalization constant. In
such a way we have a Schr\"{o}dinger cat state ${\cal N}_0\left[ |\alpha
\rangle +\exp (i\phi )|-\alpha \rangle \right] (\alpha =|\alpha |\exp
(i\theta ))$ in the limits $M\rightarrow \infty ,\eta \rightarrow 0$ with
fixed finite $\eta ^2M=|\alpha |^2$. 
Using Eq.(1), we rewrite the superposition state as
\begin{equation}
|\phi ,\eta _C,M\rangle ={\cal N}{(1-\eta ^2)}^{M/2}\sum_{n=0}^\infty 
{M+n-1 \choose n}\eta ^n_C[1+\exp (i\phi )(-1)^n]|n\rangle
\end{equation}
where 
\begin{equation}
{\cal N}=\left\{ 2\left[ 1+\cos \phi {\frac{{(1-\eta ^2)}^M}{{(1+\eta ^2)}^M}%
}\right] \right\} ^{-{\frac 12}}
\end{equation}
is the normalization constant. In deriving Eq.(4), we have used the inner
products of the NBSs

\begin{eqnarray}
\langle \alpha ,M|\beta ,M\rangle &=&(1-|\alpha |^2)^{M/2}(1-|\beta
|^2)^{M/2}(1-\alpha ^{*}\beta )^{-M}, \\
\langle -\eta _C,M|\eta _C,M\rangle &=&\langle \eta _C,M|-\eta _C,M\rangle
=(1-\eta ^2)^M(1+\eta ^2)^{-M}.  \nonumber
\end{eqnarray}

From Eqs.(3) and (4), the photon number distribution of the
superposition state is obtained as 
\begin{equation}
P(n)={M+n-1\choose n}\frac{\eta ^{2n}[1+\cos \phi (-1)^n]}{{(1-\eta
^2)^{-M}+\cos \phi (1+\eta ^2)^{-M}}}.
\end{equation}
For $\phi =0(\phi =\pi )$, we see that the odd(even) photon numbers are
excluded from $P(n)$. So these states could be nominated as even (odd)
NBS in analogy to the even(odd) coherent state and the even
(odd) binomial state. From Eq.(6), we can see the
characteristic oscillations in photon number for $\phi =0(\phi =\pi )$. For $%
\phi =\pi /2$, the photon number distribution of the superposition state becomes smooth,
and it is identical to that of the NBS $|\eta _C,M\rangle $.


\section{Non-classical properties}

The simplest way to investigate the statistical properties of the field
is to differentiate the generating function 
\begin{equation}
G(\lambda )=\sum_{n=0}^\infty P(n)\lambda ^n={\frac{{(1-\lambda \eta
^2)^{-M}+\cos \phi (1+\lambda \eta ^2)^{-M}}}{{(1-\eta ^2)^{-M}+\cos \phi
(1+\eta ^2)^{-M}}}}
\end{equation}
with respect to the auxiliary real parameter $\lambda $. From the generating
function, the mean photon number and the expectation value of the square of
photon number operator are derived as 
\begin{eqnarray}
\langle a^{\dagger }a\rangle &=&{\frac{{M\eta ^2[(1-\eta ^2)^{-M-1}-\cos
\phi (1+\eta ^2)^{-M-1}]}}{{(1-\eta ^2)^{-M}+\cos \phi (1+\eta ^2)^{-M}}}}, 
\\
\langle (a^{\dagger }a)^2\rangle &=&\frac{M\eta ^2[(1-\eta ^2)^{-M-1}-\cos
\phi (1+\eta ^2)^{-M-1}]}{(1-\eta ^2)^{-M}+\cos \phi (1+\eta ^2)^{-M}} \nonumber\\
&&+\frac{M(M+1)\eta ^4[(1-\eta ^2)^{-M-2}+\cos \phi (1+\eta ^2)^{-M-2}]}{%
(1-\eta ^2)^{-M}+\cos \phi (1+\eta ^2)^{-M}}.
\end{eqnarray}
Here $a$ and $a^\dagger$ are the annihilation and creation operators of the radiation field, respectively. When $\phi =\pi/2,$ the mean photon number becomes $\langle {a^{\dagger
}a\rangle =M\eta }^2/(1-\eta ^2),$ which is the mean photon number of the
NBS. It is proportional to the parameter $M.$

The sub-Poissonian properties of the radiation field are characterized by
Mandel's $Q$ parameter 
\begin{equation}
Q={\frac{{\langle \Delta (a^{\dagger }a)^2\rangle -\langle a^{\dagger
}a\rangle }}{{\langle a^{\dagger }a\rangle }}},
\end{equation}
where $\langle \Delta (a^{\dagger }a)^2\rangle =\langle (a^{\dagger
}a)^2\rangle -\langle a^{\dagger }a\rangle ^2$ is the fluctuation of the
photon number. When $Q<0$, $Q=0$ and $Q>0$, the states are called
sub-Poissonian, Poissonian and super-Poissonian, respectively. It is easy to
calculate that 
\begin{eqnarray}
Q &\equiv &Q(\phi ,\eta ,M)=\frac{(M+1)\eta ^2\left[ (1-\eta ^2)^{-M-2}+\cos
\phi (1+\eta ^2)^{-M-2}\right] }{(1-\eta ^2)^{-M-1}-\cos \phi (1+\eta
^2)^{-M-1}}  \nonumber \\
&&-\frac{M\eta ^2\left[ (1-\eta ^2)^{-M-1}-\cos \phi (1+\eta
^2)^{-M-1}\right] }{(1-\eta ^2)^{-M}+\cos \phi (1+\eta ^2)^{-M}}.
\end{eqnarray}
Taking into account Eq.(8) we finally find that 
\begin{equation}
Q=\langle \pi -\phi ,\eta ,M+1|N|\pi -\phi ,\eta ,M+1\rangle-\langle \phi,\eta ,M |N |\phi,\eta ,M\rangle 
\end{equation}

It is easy to see that $Q(2\pi -\phi ,\eta ,M)=Q(\phi ,\eta ,M)$, namely, $Q$
is symmetric with respect to $\phi =\pi $. So in the following we only
consider the range $0\leq \phi \leq \pi $. However, for the extreme cases $%
\phi =0$ and $\phi =\pi $ with $\eta ^2\to 0$, we have to consider them
separately since one of the denominators in Eq.(11) is vanishing
in the two extreme cases. Fortunately, in both cases, the limits exist, namely, 
\begin{equation}
\lim_{\eta ^2\to 0}Q(0,\eta ,M)=1,\ \ \ \ \ \lim_{\eta ^2\to 0}Q(\pi ,\eta
,M)=-1.
\end{equation}
For any other $\phi \neq 0,\pi $, $Q(\phi ,\eta ,M)\to 0$ in the limit $\eta
^2\to 0$. So in the limit $\eta ^2\to 0$, $Q(\phi ,\eta ,M)$ is independent
of $M$. Besides, it is not difficult to see that $Q(\phi ,\eta,M)\to \infty $
in the limit $\eta^2\to 1$ and it does not depend on $\phi$ . We shall analyse this fact later.

In figure 1 we plot the $Q$ parameter against $\eta $ for different $\phi $. For
the case $\phi =\pi /2$, the superposition state is super-Poissonian just as
the NBS because of the identical photon number distributions of the two
states. This fact can also be seen from Eq.(12). When $\phi =\pi /2,$
Eq.(12) becomes $Q=\langle \eta,M+1|N{|\eta ,M+1\rangle }-\langle\eta,M|N{|\eta ,M\rangle }.$ Since the mean photon number increases with the increase of $%
M$,$Q$ is always larger than 0 and the NBS is super-Poissonian. As
seen from the figure the even NBS ($\phi =0$) is always super-Poissonian and
the odd NBS ($\phi =\pi $) can be sub-Poissonian when $\eta $ is small. For
the case with $\phi =3\pi /4$ the superposition state can be sub-Poissonain
for small $\eta $. But the sub-Poissonian range and degree are less than
those of the odd NBS. It is an interesting feature that, when $\eta $ is large
enough, the four curves merge to one curve as seen from the figure, and the
states are all super-Poissonian independent of $\phi $. In fact, this is
because $(1-\eta ^2)^{-M}\gg (1+\eta ^2)^{-M}$ for large enough $\eta ^2$.
For the smallest $M=1$, $(1-\eta ^2)^{-1}/(1+\eta ^2)^{-1}=19$ for $\eta
^2=0.9$. Of course,
the larger $M$, the larger $(1-\eta ^2)^{-M}/(1+\eta ^2)^{-M}$.

In order to discuss the squeezing effects of the superposition states of the NBSs,
we first calculate the expectation value of operator $a^k$ from Eq.(2) 
\begin{eqnarray}
\langle a^k\rangle &=&{\cal N}^2\sqrt{\frac{{(M+k-1)!\eta ^{2k}}}{{%
(M-1)!(1-\eta ^2)^k}}}\{ A_{k+}[1+(-1)^k]+  \nonumber \\
&&A_{k-}[\cos \phi (1+(-1)^k)-i\sin \phi (1-(-1)^k)]\}\exp (ik\theta ),
\end{eqnarray}
where 
\begin{equation}
A_{k\pm }=(1-\eta ^2)^{M+k/2}\sum_{n=0}^\infty {{M+n-1}\choose{n}}^{\frac 12}{%
{M+n+k-1}\choose{n}}^{\frac 12}(\pm \eta ^2)^n.
\end{equation}
Eq.(14) can be reduced to the following equations for the cases $%
k=1,\ 2$, 
\begin{eqnarray}
\langle a\rangle &=&{\frac{{2{\cal N}^2\sqrt{M}\eta \sin \phi A_{1-}}}{{%
\sqrt{1-\eta ^2}}}}(\sin \theta -i\cos \theta )  \nonumber \\
\langle a^2\rangle &=&{\frac{{2{\cal N}^2\sqrt{M(M+1)}\eta ^2}}{{1-\eta ^2}}}%
(A_{2+}+A_{2-}\cos \phi )\exp (i2\theta )
\end{eqnarray}

Define the quadrature operators $X_1=(a+a^\dagger)/2, X_2=(a-a^\dagger)/(2i)$%
. Then their variances are 
\begin{eqnarray}
\langle\Delta(X_1)^2\rangle &=& {\frac{1}{4}}+{\frac{1}{2}}\left[\langle a^+
a\rangle + \mbox{Re}\langle a^2 \rangle -2(\mbox{Re}\langle a \rangle)^2
\right],  \nonumber \\
\langle\Delta(X_2)^2\rangle &=& {\frac{1}{4}}+{\frac{1}{2}}[\langle a^+
a\rangle  -\mbox{Re}\langle a^2 \rangle -2(\mbox{Im}\langle a \rangle)^2].
\end{eqnarray}
The field is considered to be squeezed when any of the varainces is below the
vacuum level, that is, either $\langle\Delta(X_1)^2\rangle<1/4 $ or $%
\langle\Delta(X_2)^2\rangle<1/4 $.

Figure 2 is a plot showing how the $\langle \Delta (X_2)^2\rangle $ depends on 
$\eta $ for different $\phi $. The odd NBS shows no
squeezing and the squeezing can occur for the cases with $\phi =0,\pi /2,3\pi
/4 $ when $\eta $ is small,. When $\eta $ increases,the squeezing effects of the superpostion
states do not depend on $\phi $ and they first show squeezing for a wide range of the parameter $\eta$ and then no
squeezing as seen from the figure. 

\section{Algebraic structures and\\
ladder operator formalisms of
the even and odd
NBSs}

The NBS admits ladder operator formalism, namely, it
satisfies the eigenvalue equation of generators of su(1,1) Lie algebra. We show that the even and odd NBSs also admit ladder operator formalism.

We consider two general states defined in even and odd Fock space,
\begin{eqnarray}
|\eta _C,M\rangle _e =\sum_{n=0}^\infty C_e(n)|2n\rangle
\\
|\eta _C,M\rangle _o =\sum_{n=0}^\infty C_o(n)|2n+1\rangle
\end{eqnarray}
 
In order to find the ladder operator formalism of the even state $|\eta _C,M\rangle _e$, we suppose that it satisfies the following eigenvalue equation
\begin{equation}
[N-f(N)a^{\dagger 2}]|\eta_C,M\rangle_e=0,
\end{equation}
in which $f(N)$ is a real function of the number operator $N$.
Substituting Eq.(18) into Eq.(20), we determine the function $f(N)$ as
\begin{equation}
f(N)=\sqrt{\frac{N}{N-1}}\frac{C_e(N/2)}{C_e(N/2-1)}.
\end{equation}
Therefore, we have obtained the ladder operator formalism of a general even state defined in the even Fock space.

Let us examine the algebraic structure involved in the general even state. Define 
${\cal A}$ as an associate algebra with generators 
\begin{equation}
{N},A_{+}=f(N)a^{\dagger 2},A_{-}=(A_{+})^{\dagger }.
\end{equation}
Then it is easy to verify that these operators satisfy the following
relations 
\begin{equation}
[{N},A_{\pm }]=\pm A_{\pm },A_{+}A_{-}=S({N}),A_{-}A_{+}=S({N}+1),
\end{equation}
where the  function 
\begin{equation}
S({N})=N^2\frac{C_e^2(N/2)}{ C_e^2(N/2-1)}.
\end{equation}

This algebra ${\cal A}$ is nothing but the generally deformed oscillator(GDO)[25]%
algebra with the structure function $S({N})$. From Eq.(3) for the case of $\phi=0$, we see that
the even NBS bears the generally deformed oscillator algebraic structure with structure function $N(M+N-1)(M+N-2)\eta_C^4/(N-1)$.

Acting the operator $a^2$ on Eq.(20) from left and using Eq.(21), we get 
\begin{equation}
a^2|\eta_C,M\rangle_e=\sqrt{(N+1)(N+2)}\frac{C_e(N/2+1)}{C_e(N/2)}|\eta_C,M\rangle_e.
\end{equation}
In the derivation of the above equation, we have used the fact that the operator $(N+2)$ is non-zero in the whole Fock space.

Now we consider the general odd state $|\eta_C,M\rangle_o$. We assume that it satisfies the equation
\begin{equation}
[(N-1)-f(N)a^{\dagger 2}]|\eta_C,M\rangle_o=0.
\end{equation}
The function $f(N)$ is determined as
\begin{equation}
f(N)=\sqrt{\frac{N-1}{N}}\frac{C_o(\frac{N-1}{2})}{C_o(\frac{N-3}{2})}.
\end{equation}

By acting the operator $a^2$ on Eq.(26) from left, we get 
\begin{equation}
a^2|\eta_C,M\rangle_o=\sqrt{(N+1)(N+2)}\frac{C_o(\frac{N+1}{2})}{C_o(\frac{N-1}{2})}|\eta_C,M\rangle_o.
\end{equation}

As seen from Eq.(26), the algebra involved in the general odd state
a generally deformed osillator algebra with structure function $S(N)= (N-1)^2 {C_o^2(\frac{N-1}{2})}/{C_o^2(\frac{N-3}{2})}$.

Now we first consider simple cases of the two general states, the even and odd coherent states, which are defined as[11-15]
\begin{eqnarray}
|\alpha \rangle _{ECS} &=&1/\sqrt{\cosh |\alpha |^2}\sum_{n=0}^\infty \alpha
^{2n}/\sqrt{(2n)!}|2n\rangle . \\
|\alpha \rangle _{OCS} &=&1/\sqrt{\sinh |\alpha |^2}\sum_{n=0}^\infty \alpha
^{2n+1}/\sqrt{(2n+1)!}||2n+1\rangle. 
\end{eqnarray}

The ladder operator formalisms can be easily obtained from Eqs(25) and (28)
\begin{eqnarray}
a^2|\alpha \rangle _{ECS} &=&\alpha ^2|\alpha \rangle _{ECS}, \\
a^2|\alpha \rangle _{OCS} &=&\alpha ^2|\alpha \rangle _{OCS}.  
\end{eqnarray}
This is just what we expected.

From Eqs.(3) and (4), the expansion of the even and odd NBSs are obtained as
\begin{eqnarray}
|\eta _C,M\rangle _e =\sqrt{\frac 2{(1-\eta ^2)^{-M}+(1+\eta ^2)^{-M}}}%
\sum_{n=0}^\infty 
{M+2n-1 \choose 2n}^{1/2}\eta _C^{2n}|2n\rangle , \\
|\eta _C,M\rangle _o \sqrt{\frac 2{(1-\eta ^2)^{-M}-(1+\eta ^2)^{-M}}}%
\sum_{n=0}^\infty  
{M+2n \choose 2n+1}
^{1/2}\eta _C^{2n+1}|2n+1\rangle .  
\end{eqnarray}
Substituting the coefficients of the even and odd NBS to Eq.(25) and (28), respectively, we obtain 
\begin{eqnarray}
a^2|\eta_C,M\rangle_e={\sqrt{(M+N)(M+N+1)}}\eta_C^2|\eta_C,M\rangle_e,\\
a^2|\eta_C,M\rangle_o={\sqrt{(M+N)(M+N+1)}}\eta_C^2|\eta_C,M\rangle_o.
\end{eqnarray}
This is the ladder operator formalism of the even and odd NBSs.

Since the operator $\sqrt{(M+N)(M+N+1)}$ is non-zero in the whole Fock space, we get
\begin{eqnarray}
1/\sqrt{(M+N)(M+N+1)}a^2|\eta_C,M\rangle_e=\eta_C^2|\eta_C,M\rangle_e,\\
1/\sqrt{(M+N)(M+N+1)}a^2|\eta_C,M\rangle_o=\eta_C^2|\eta_C,M\rangle_o.
\end{eqnarray}
It can be seen that the even and odd NBSs satisfy a same eigenvalue equation with a same eigenvalue $\eta_C^2$.

In analogy to the definition of the one-photon nonlinear coherent states[26-27], we can naturally define the two-photon nonlinear coherent state as
\begin{equation}
F(N)a^2|\alpha\rangle=\alpha|\alpha\rangle.
\end{equation}
By this definition, we see that the even and odd NBSs can be viewed as two-photon nonlinear coherent states.

\section{Generation of the superposition states}


In this section, we discuss how to generate such superposition states of the NBSs.
One scheme involving a Kerr medium could be utilized to generate such
states with relative angle $\phi={\frac{ \pi }{2}}$ in Eq.(2). The injection of a
prepared field state into a nonlinear Kerr medium is described by the
following Hamiltonian [28,29] 
\begin{equation}
{H}_1=\hbar g_1 (a^\dagger a)^2.
\end{equation}
where $g_1$ is the nonlinear parameter of the Kerr medium. The
corresponding unitary operator is $U_1=\exp[-ig_1(a^\dagger a)^2 t]$. At time 
$t={{\pi}/{2g_1}}$,it becomes 
\begin{equation}
U_1({\frac{\pi}{2g_1}})={\frac{1}{\sqrt{2}}}[ \exp(-i{\frac{\pi}{4}}%
)+(-1)^{a^\dagger a} \exp(i{\frac{\pi}{4}})].
\end{equation}

If the initial state is prepared in a NBS $|\eta_c,M\rangle^-$. From Eq.(1)
and Eq.(41),the state vector at time $t={{\pi}/{2g_1}}$ is 
\begin{equation}
\left|\psi_1\left({\frac{\pi}{2g_1}}\right)\right\rangle= U_1\left({\frac{\pi%
}{2g_1}}\right)|\eta_C,M\rangle ={\frac{1}{\sqrt{2}}}\exp\left(-i{\frac{\pi%
}{4}}\right) \left[|\eta_C,M\rangle +\exp\left(i{\frac{\pi}{2}}
\right)|-\eta_C,M\rangle \right].
\end{equation}
The above state is just the superpostion state of the NBSs with relative
angle $\phi={\frac{\pi}{2}}$.

Next we show how to generate the superpostion states of the NBSs for arbitrary $\phi$.
We begin by assuming that we have a cavity in which the NBS has already been
prepared. The NBS can be prepared by the method of state reduction in
optical processes [8] and in the non-degenerate parameter amplifier [9]. We assume
that the atom is tuned to have a dispersive interaction with the cavity and
the atom does not exchange energy with the cavity field. In the interaction
picture, the effective Hamiltonian for the non-resonant atom-field
interaction is given by[30] 
\begin{equation}
H_2=\hbar g_2 a^\dagger a \sigma_3,
\end{equation}
where $\sigma_3$ is the atomic inversion operator and $g_2$ is a parameter
depending on the dipole moment and detuning between the atomic transition
frequency and the cavity frequency.

The atom can be prepared in the superposition state ${\frac{1}{\sqrt{2}}}%
[|g\rangle + \exp(i\phi)|e\rangle]$. The exicited state of the atom is
denoted by $|e\rangle$ and ground state by $|g\rangle$. The initial state of
the composite system can be written as 
\begin{equation}
|\psi_2(0)\rangle={\frac{1}{\sqrt{2}}}[|g\rangle+
\exp(i\phi)|e\rangle]\otimes| \eta_C,M\rangle
\end{equation}
The state vector at time $t$ is easily obtained as 
\begin{eqnarray}
|\psi_2(t)\rangle &=&\exp(-iH_2 t)|\psi_2(0)\rangle  \nonumber \\
&=& {\frac{1}{\sqrt{2}}}\left[|g\rangle\otimes |\eta_C,M\rangle
+\exp(i\phi)|e\rangle\otimes |\eta_C \exp(-ig_2t),M\rangle \right].
\end{eqnarray}

Then, the second classical microwave field induces a $\pi$ pulse on the
atom, causing the transitions $|g\rangle\rightarrow (|g\rangle-|e\rangle)/%
\sqrt{2},\ |e\rangle\rightarrow(|g\rangle+|e\rangle)/\sqrt{2}$. After this
manipulation, we obtain the state 
\begin{eqnarray}
|\psi^{\prime}_2(t)\rangle &=& {\frac{1}{2}}\left\{
|g\rangle\otimes|\eta_C,M\rangle +
\exp(i\phi)|\eta_C\exp(-ig_2t),M\rangle] \right.  \nonumber \\
& & \left. +|e\rangle\otimes \left[\exp(i\phi)
|\eta_C\exp(-ig_2t),M\rangle -|\eta_C,M\rangle \right]\right\}.
\end{eqnarray}

Finally the atom is selected ionized. If we find the atom in the state $%
|g\rangle$ and let $g_2t=\pi$ with appropriate atomic velocity choices, the
cavity field is projected into the state 
\begin{equation}
\left|\psi_2^{\prime}\left({\pi/g_2}\right)\right\rangle ={\cal N}\left[|\eta_C,M%
\rangle +\exp(i\phi)|-\eta_C,M\rangle\right],
\end{equation}
The above states are just the
superposition states of the NBSs. Therefore, the superposition states for arbitrary relative
angle $\phi$ are obtained.


\section{Summary}

We have introduced the superpositions of the NBSs in such a way
that they can be reduced to Schr\"odinger cat states in a certain limit.
Their photon number distributions can show oscillatory behaviors. It is well known
that the NBS are super-Poissonian, but the superpositions of the NBSs can be
sub-Poissonian for small $\eta$. For large $\eta$, the Mandel's Q parameter
does not depend on the relative phase $\phi$ and the states are
super-Poissonian. The superposition states can exhibit squeezing and the
squeezing does not depend on related phase for large $\eta$ just as the behaviors of the Q
parameter. 
The ladder operator formalisms have been obtained for the two general states defined in the even and odd Fock space. These two general states bear GDO algebraic structures with different structure functions. As special cases of the two general states, even and odd NBSs,
also admit ladder operator formalism and bear GDO algebraic structure. It is found that the even and odd NBSs satisfy a same eigenvalue equation with a same eigenvalue and they can be viewed as two-photon nonlinear coherent states.
We proposed two methods to genertate the superposition states of
the NBSs. The superposition states with relative phase $\pi/2$ can be generated
in nonlinear Kerr medium and those with arbitrary relative phase $\phi$ can
be generated in the context of cavity quantum electrodynamics.

\vspace{1cm}

{\bf Acknowledgement::}\\
This work is supported in part 
by the National 
Science Foundation of China with grant number:19875008.
One of the authors, X.-G. Wang thanks for the help of Profs.C.P.Sun,
Shao-Hua Pan and Guo-Zhen Yang.

\vspace{1cm}

{\bf Figure Captions:}\\
Fig.1, The $Q$ parameter against $\eta$ for different $\phi$. The parameter $M=30$.\\
Fig.2, The variance $\langle\Delta(X_2)^2\rangle$ against $\eta$ for different $\phi$. The parameter $M=50$.

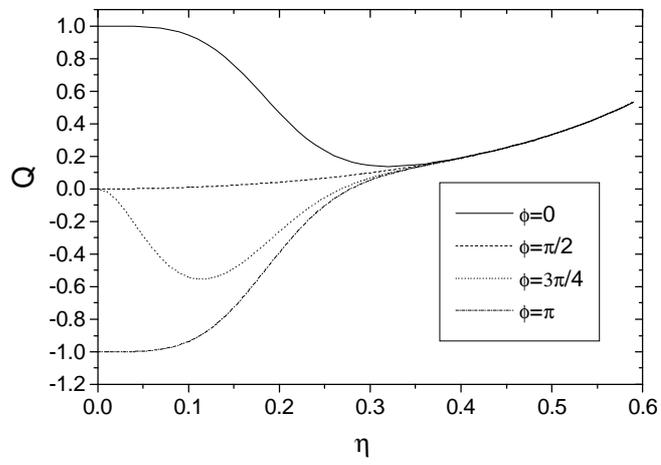
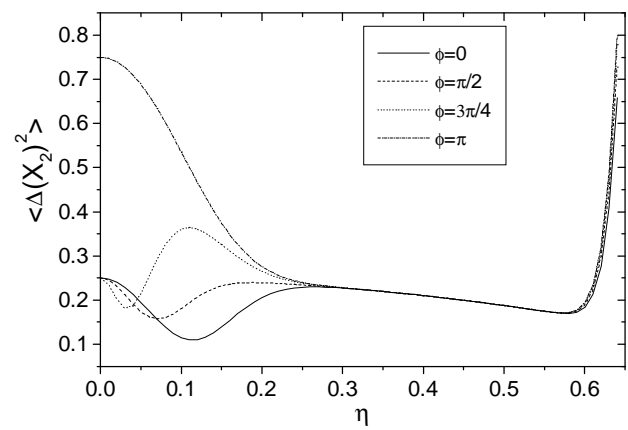
\end{document}